\documentstyle[aps,multicol,prl,array]{revtex}
\begin{document}
\draft
\title{Inequivalence of pure state ensembles for open quantum 
systems: \\
the preferred ensembles are those that are physically realizable}
\author{H.M. Wiseman$^{1,2}$, John A. Vaccaro$^{2}$}
\address{$^{1}$Centre for Quantum Dynamics, School of Science, Griffith University,  Brisbane,
Queensland 4111 Australia. \\
$^{2}$Division of Physics and Astronomy, University of
Hertfordshire, Hatfield AL10 9AB, UK.}
\maketitle

\newcommand{\beq}{\begin{equation}}
\newcommand{\eeq}{\end{equation}}
\newcommand{\bqa}{\begin{eqnarray}}
\newcommand{\eqa}{\end{eqnarray}}
\newcommand{\nn}{\nonumber}
\newcommand{\nl}[1]{\nn \\ && {#1}\,}
\newcommand{\erf}[1]{Eq.~(\ref{#1})}
\newcommand{\dg}{^\dagger}
\newcommand{\rt}[1]{\sqrt{#1}\,}
\newcommand{\smallfrac}[2]{\mbox{$\frac{#1}{#2}$}}
\newcommand{\half}{\smallfrac{1}{2}}
\newcommand{\bra}[1]{\langle{#1}|}
\newcommand{\ket}[1]{|{#1}\rangle}
\newcommand{\ip}[2]{\langle{#1}|{#2}\rangle}
\newcommand{\sch}{Schr\"odinger }
\newcommand{\schs}{Schr\"odinger's }
\newcommand{\hei}{Heisenberg }
\newcommand{\heis}{Heisenberg's }
\newcommand{\bl}{{\bigl(}}
\newcommand{\br}{{\bigr)}}
\newcommand{\ito}{It\^o }
\newcommand{\str}{Stratonovich }
\newcommand{\dbd}[1]{{\partial}/{\partial {#1}}}
\newcommand{\sq}[1]{\left[ {#1} \right]}
\newcommand{\cu}[1]{\left\{ {#1} \right\}}
\newcommand{\ro}[1]{\left( {#1} \right)}
\newcommand{\an}[1]{\left\langle{#1}\right\rangle}
\newcommand{\implies}{\Longrightarrow}
\newcommand{\tr}[1]{{\rm Tr}\sq{ {#1} }}
\newcommand{\del}{\nabla}
\newcommand{\du}{\partial}
\renewcommand{\section}[1]{{\em #1}.}

\begin{abstract}
An open quantum system in steady state $\hat\rho_{\rm ss}$ can be
represented by a weighted ensemble of pure states $\hat\rho_{\rm
ss}=\sum_{k}\wp_{k}\ket{\psi_k}\bra{\psi_k}$ in infinitely many
ways. A physically realizable (PR) ensemble is one for which some
continuous measurement of the environment will collapse the system
into a pure state $\ket{\psi(t)}$, stochastically evolving such
that the proportion of time for which $\ket{\psi(t)} =
\ket{\psi_{k}}$ equals $\wp_{k}$.  Some, but not all, ensembles are PR.
This constitutes the {\em preferred
ensemble fact}. 
 We present the necessary and sufficient conditions
for a given ensemble to be PR, and illustrate the method by
showing that the coherent state ensemble is {\em not} PR for an atom
laser.  
\end{abstract}


\begin{multicols}{2}
\narrowtext

\section{Introduction}
It is well known that there are infinitely many ways to write a
given mixed state as a mixture of (in general {\em
non-orthogonal}) pure states. That is, there are infinitely many
ensembles, $\{(\hat{\Pi}_{k},\wp_{k})\}_{k}$, consisting of
ordered pairs of pure states (rank-one projectors
$\hat\Pi_{k}=\ket{\psi_{k}}\bra{\psi_{k}}$) and weights $\wp_k$,
that represent a given state matrix (density operator) $\hat\rho$
according to
\beq
\hat\rho = \sum_{k} \wp_k\hat\Pi_{k}.
\eeq
Any such
ensemble representation suggests an {\em ignorance interpretation}
of the state matrix $\hat\rho$. That is, one would claim that the
system ``really'' is in one of the pure states $\hat\Pi_{k}$, but
that one happens to be ignorant of which state $\hat\Pi_{k}$ (i.e.
which $k$) pertains. The weight $\wp_k$ is interpreted as the
probability that the system has state $\hat\Pi_{k}$.

This interpretation can only be maintained
 for so-called proper mixtures, i.e. those for which the system
 is not entangled with its environment \cite{dEs89}.
However, an improper mixture may be turned into a proper mixture
simply by appropriately measuring the environment, and ignoring
the result. The {\em partition ensemble fallacy} \cite{KokBra00}
or {\em preferred ensemble fallacy} is that one may use a
particular ensemble to draw inferences about an experiment, even
if those inferences depend upon the choice of ensemble. Here we do
not deny that this is indeed a fallacy.

Mixed states arise naturally in the description of {\em open
quantum systems}, in fields as diverse as chemical physics,
quantum optics, micromechanics and nano-electronics. It is often
possible to approximate the evolution of such systems by a
Markovian master equation $\dot{\hat\rho}={\cal L}\hat\rho$, where
${\cal L}$ is the Lindbladian \cite{GarZol00}. Under these
conditions, an  experimenter may perform continual measurements on
the environment with which the system interacts without affecting
the master equation evolution \cite{GarZol00,fn1}. Hence the
system state can be considered a proper mixture.

Let us restrict the
discussion to a stationary 
mixed state $\hat\rho_{\rm ss}$, assumed unique:
\beq \label{ME}
  \dot{\hat\rho}_{\rm ss} = {\cal L}\hat\rho_{\rm ss} = 0.
\eeq
Also, let us consider only {\em stationary  ensembles} for
$\hat\rho_{\rm ss}$. Clearly, once the system has reached
steady-state then such a stationary ensemble will represent the
system {\em for all times} $t$. Now, if the ignorance
interpretation were to hold for such an ensemble then it should be
possible, in principle, for the experimenter to know which state
$\hat\Pi_{k}$ pertains to the system at any particular time $t$.
The pertinent index $k$ would change stochastically such that the
proportion of time the system has state $\hat\Pi_{k}$ is $\wp_k$.

In this letter we show that for some ensembles this ignorance
interpretation fails. That is, there is no way that an
experimenter can know at all (long) times that the system has some
ensemble state $\hat\Pi_{k}$. We say that such ensembles are not
{\em physically realizable} (PR). However, there are other
stationary ensembles that {\em are} PR. 

The existence, for a given system, of two non-empty
classes of stationary ensembles, those that are PR and those that
are not, constitutes a {\em preferred ensemble fact} (PE-fact).
Note that this fact identifies a preferred class of ensembles (the
PR ones), not a unique preferred ensemble. Also note that this
fact does not contradict the argument against preferred ensembles
in Ref.~\cite{KokBra00}, because it is a fact about ensembles
representing a stationary state {\em for all times}. As we will
see, the PE-fact has surprising implications for some open
quantum systems.

This letter is organized as follows. First we
discuss the Hughston-Josza-Wootters theorem \cite{HugJozWoo93} and
why it does not contradict the PE-fact. Then we give the necessary
and sufficient conditions for an ensemble to be PR for a given
system. For linear systems with uniform Gaussian ensembles we show
these conditions reduce to a simple inequality. Next we establish
the PE-fact for a particular system of interest: a model atom
laser. Finally we discuss the implications of our results.

\section{The HJW Theorem}
We wish to consider a system with state matrix $\hat\rho$ which is
mixed solely due to its entanglement with a second system, the
{\em environment}. That is, there is always a pure state
$\ket{\Psi}$ in a larger Hilbert space of system plus environment
such that
\beq \label{rhored}
\hat\rho = {\rm Tr}_{\rm env}\sq{\ket\Psi\bra\Psi}.
\eeq
Certainly it is always possible formally to find such a pure state
\cite{HugJozWoo93}.
Also, if the environment is initially not pure then it can be then
measured in its diagonal basis so as to make it conditionally pure
without affecting its subsequent interaction with the system.
Therefore we can assume the existence of $\ket{\Psi}$ without loss
of generality.

The  Hughston-Jozsa-Wootters (HJW) theorem \cite{HugJozWoo93} says
the following. For any ensemble $\{(\hat\Pi_{k},\wp_k)\}_{k}$ that
represents $\hat\rho$, it is possible to measure the environment
such that the system state is collapsed into one of the pure
states $\hat\Pi_{k}$ with probability $\wp_k$. That is, there
exists a probability-operator measure $\{\hat{F}_{k}\}_{k}$ acting
on the {\em environment} Hilbert space such that
\beq
\wp_k{\hat\Pi}_{k} = {\rm Tr}_{\rm env}\sq{
{\hat{1}_{\rm sys}\otimes \hat{F}_{k}^{1/2}}
 \ket{\Psi}\bra{\Psi} \hat{F}_{k}^{1/2}\otimes
{\hat{1}_{\rm sys}}}.
\eeq

The HJW theorem gives rigorous meaning to the ignorance
interpretation of any particular ensemble. It says that there will
be a way to perform a measurement on the environment, without on
average disturbing the system state, to obtain exactly the
information as to which state the system is ``really'' in. Of
course the fact that one can do this for any ensemble means that
no ensemble can be fundamentally preferred over any other one, as
a representation of $\hat\rho$ {\em at some particular time $t$}.

The HJW theorem does not contradict the preferred ensemble fact
introduced above. This is because the PE-fact refers to the much
stronger notion of representing the state of the system obeying a
master equation, at steady state, by a {\em stationary} ensemble
that applies {\em at all times}. This extra condition means that
the HJW theorem does not apply, and not all ensembles are PR
through measuring the environment of the system. We will now give
the conditions for an ensemble to be PR.

\section{Establishing the Preferred Ensemble Fact}
Consider a system obeying the master equation
$\dot{\hat\rho}={\cal L}\hat\rho$ in the long-time limit so that
it is described by the (assumed unique) stationary state
$\hat\rho_{\rm ss}$. Now say we wish to give an ignorance
interpretation at all long times to a particular stationary
ensemble $\{(\hat{\Pi}_{k},\wp_{k})\}_{k}$ satisfying
\beq \label{cond0}
\hat\rho_{\rm ss} = \sum_{k} \wp_{k}\hat\Pi_{k}.
\eeq
 At a particular time $t$ this is always possible by the HJW
theorem. That is, we can measure the environment to find out the
pertinent state $\hat\Pi_{k}$, and, on average, the system remains
in state $\hat\rho_{\rm ss}$. This may involve measuring parts of
the environment that interacted with the system an arbitrarily
long time ago, but there is nothing physically impossible in doing
this.

Now consider the  future evolution of the system state following
this measurement. At time $t+\tau$, it
will have evolved 
to $\hat\rho_{k}(t+\tau)=\exp({\cal L}\tau)\hat\Pi_{k}$.
This is a mixed state because the system has now become re-entangled
with its environment.
The system state can be repurified by making another measurement
on its environment. However, if the same ensemble is to remain as
our representation of the system state then the pure system states
obtained as a result of this measurement at time $t+\tau$ must be
contained in $\{\hat\Pi_{k}\}_{k}$. By the HJW theorem, this will
possible if and only if  $\hat\rho_{k}(t+\tau)$ can be represented
as a mixture of these states. That is, for all $k$ there must
exist a probability measure $\cu{w_{jk}(\tau)}_{j}$ such that
\beq \label{cond1}
\exp({\cal L}\tau)\hat\Pi_{k} = \sum_{j} w_{jk}(\tau) \hat\Pi_{j}.
\eeq
If $w_{jk}(\tau)$ exists then it is the probability
that the measurement at time $t+\tau$  yields the state $\hat\Pi_{j}$.

Equation (\ref{cond1}) is necessary but not sufficient for the ensemble
$\{(\hat\Pi_{j},\wp_{j})\}_{j}$ to be PR.
We also require that the  weights
be stationary. That is, that for all
$j$ and all $\tau$,
\beq \label{cond2}
\wp_{j} = \sum_{k} \wp_{k}w_{jk}(\tau).
\eeq
Multiplying both sides of \erf{cond1} by $\wp_{k}$, and summing
over $k$, then using \erf{cond2} and \erf{cond0} gives $ e^{{\cal
L}\tau} \hat\rho_{\rm ss} = \hat\rho_{\rm ss}$, as required from
the definition of $\hat\rho_{\rm ss}$.

If the two conditions (\ref{cond1}) and (\ref{cond2}) are
satisfied for some time $\tau$ then they will be satisfied for any
multiple of $\tau$. Therefore it is sufficient to establish that
they are satisfied for $\tau=dt$. (Here $dt$ is infinitesimal with
respect to all relevant system time scales but strictly must be
long compared to the environment correlation time; the master
equation is not valid on any shorter time scale.) For a bounded
Lindbladian superoperator ${\cal L}$ and distinct states
$\{\hat\Pi_{k}\}_{k}$, if $w_{jk}(dt)$ exists then
it is given by
\beq
  w_{jk}(dt) = \delta_{kj}- \delta_{kj}\sum_{l}\gamma_{kl}dt + \gamma_{kj}dt
  \label{w_eqn}
\eeq
with $\gamma_{kj}$ finite. The coefficient $\gamma_{kj}$ may be
interpreted as the rate for the system to jump from state
$\hat\Pi_{k}$ to state $\hat\Pi_{j}$. That is, the quantum master
equation is replaced by a classical master equation \cite{Gar85}
for the occupation probabilities $\cu{p_{k}}_{k}$ for the states
$\{\hat\Pi_{k}\}_{k}$ where $p_j(t+dt)=\sum_{k} p_{k}(t)
w_{jk}(dt)$, and so from \erf{w_eqn}
\beq \label{clasme}
\dot{p}_{k} = -p_{k}\sum_{j}\gamma_{kj} + \sum_{j}\gamma_{kj}p_{j}.
\eeq
The condition (\ref{cond2}) is then
equivalent to the condition that $p_{k}=\wp_{k}$ be the stationary
solution of \erf{clasme}.

Any stationary scheme of continual rank-one measurements of the
environment will produce stochastic dynamics of this sort in the
steady state \cite{WisVac98}. The ignorance interpretation of the
ensemble thus produced is then rigorously justified. The PE-fact
follows if, for the given system, there exists at least one
ensemble satisfying \erf{cond0} for which there does not exist a
probability measure $\cu{w_{jk}(\tau)}_{j}$ satisfying conditions
(\ref{cond1}) and (\ref{cond2}).

The replacement of a quantum master equation by a classical master
equation where the classical index $k$ is associated with a
quantum state $\hat\Pi_{k}$ is often used to provide an intuitive
picture of irreversible quantum dynamics. A canonical example is
Einstein's description \cite{Ein17} of an atom driven by a thermal
field in terms of jumps from ground to excited states (absorption)
and from excited states to ground (emission). Of course Einstein
did not know the more general description in terms of the quantum
master equation, but it is easy to verify that the the ensemble
consisting of atomic energy eigenstates is a PR ensemble.
Specifically, to realize this ensemble one must continually
measure the environment in the photon number basis.

\section{Linear Dynamics and Uniform Gaussian Ensembles}
The description given above applies most naturally to ensembles
with a discrete set of states $\{\hat\Pi_{k}\}_{k}$. In many cases
we wish to consider a continuum of states. In these cases it is
convenient to take a limit in which the jump process described by
rates $\gamma_{kj}$ is replaced by a diffusion process. We
restrict ourselves to systems with linear dynamics, and uniform
Gaussian ensembles. These terms (defined below) only apply to
quantum systems whose state $\hat\rho$ can be represented by a
Wigner function \cite{GarZol00} on $2n$-dimensional Euclidean
phase space:
\beq
W(\vec{z}) =  \tr{\hat\rho  \; S\prod_{m=1}^{2n}
\int \frac{d\xi_{m}}{2\pi}\, \exp\sq{i\xi_{m}(z_{m}-\hat{z}_{m})}   }.
\eeq
Here $S$ denotes ordering  symmetric in the operators 
\beq
(\hat{z}_{1},\cdots \hat{z}_{2n})
 =
(\hat{x}_{1},\hat{p}_{1},\hat{x}_{2},\hat{p}_{2},\cdots,\hat{x}_{n},\hat{p}_{n})
,
\eeq
where the co-ordinates $\cu{\hat{x}_{n}}_{n}$ and momenta 
$\cu{\hat{p}_{n}}_{n}$ each form a mutually
commuting set of operators with the reals as eigenvalues, but (with $\hbar=2$)
$[\hat{x}_{n},\hat{p}_{n'}]=2i\delta_{n,n'}$.

Such a system has linear dynamics if and only if (for a suitable
choice of co-ordinates and momenta) its Wigner function obeys an
Ornstein-Uhlenbeck equation \cite{Gar85}
\beq \label{OUE}
\dot{W}(\vec{z}) = \ro{ \vec{\del}^{T}{\bf K} \vec{z} + \half
\vec{\del}^{T} {\bf D} \vec{\del} } W(\vec{z}).
\eeq
Here $\del_{m} = \dbd{z_{m}}$, ${\bf K}$ is a constant matrix, and
${\bf D}$ is a constant matrix that is symmetric and positive
semidefinite. For simplicity we will assume that the eigenvalues of 
${\bf K}$ have positive real parts, so that the system has a stationary state
\beq
W_{\rm ss}(\vec{z}) = G(\vec{z};\vec{0},{\bf V}_{\rm ss}),
\eeq
where ${\bf V}_{\rm ss}$ is defined by \cite{Gar85}
\beq \label{defVinf}
{\bf K}{\bf V}_{\rm ss}+{\bf V}_{\rm ss}{\bf K}^{T} = {\bf D}.
\eeq
Here we are using the notation that $G(\vec{z};\vec{\mu},{\bf V})$
is a multivariate Gaussian in $\vec{z}$ parameterized by the mean
vector $\vec{\mu}$ and the covariance matrix ${\bf V}$
\cite{Gar85}.

A uniform Gaussian ensemble of pure states
$\{(\hat{\Pi}_{k},\wp_{k})\}_{k}$ comprises states $\hat\Pi_{k}$
that have Wigner functions $W_{k}(\vec{z})$ that are Gaussians
\beq
W_{k}(\vec{z}) \equiv W(\vec{z};\vec{\mu}_{k}) =
G({\vec z};\vec{\mu}_{k},{\bf V})
\eeq
with mean determined by $k$ but variance independent of $k$. That
is, the states all have the same ``shape'', but have different
displacements in $2n$-dimensional phase space. Since the ensemble
represents $\hat\rho_{\rm ss}$ we have (converting from the
discrete $k$ to the continuum variable $\vec{\mu}$)
\beq
W_{\rm ss}(\vec{z}) = \int d\mu_{1} \cdots \int d\mu_{2n}\, \wp(\vec{\mu})
W(\vec{z};\vec{\mu}).
\eeq
Since $W_{\rm ss}(\vec{z})$ and $W(\vec{z};\vec{\mu})$ are
Gaussians, $\wp(\vec{\mu})$ is also:
\beq \label{defwp}
\wp(\vec{\mu}) = G(\vec{\mu};\vec{0},{\bf V}_{\rm ss} - {\bf V}).
\eeq

Now consider the conditions for the uniform Gaussian ensemble to
be PR, starting with \erf{cond1}. From \erf{OUE}, if the system
begins with Wigner function $W(\vec{z};\vec{\mu})$ then after an
infinitesimal time $dt$ its Wigner function will be
\beq
G(\vec{z};\vec{\mu} - {\bf K}\vec{\mu}dt,{\bf V} +  {\bf D}dt
- {\bf K}{\bf V}dt - {\bf V}{\bf K}^{T}dt).
\eeq
This can be written as a mixture of the ensemble states
\beq
\int d\vec{\mu}' w(\vec{\mu}';\vec{\mu};dt) W(\vec{z};\vec{\mu})
\eeq
if and only if the transition probability density is
\beq \label{transprob}
w(\vec{\mu}';\vec{\mu};dt) = G(\vec{\mu}';\vec{\mu}-{\bf K}\vec{\mu}dt,
{\bf D}dt - {\bf K}{\bf
V}dt - {\bf V}{\bf K}^{T}dt).
\eeq
Since the ensemble is specified completely by ${\bf V}$,
\erf{cond1} turns into the condition on {\bf V} that
\beq \label{defB}
{\bf B}_{\bf V} \equiv {\bf D}- {\bf K}{\bf V} - {\bf V}{\bf K}^{T} \geq 0.
\eeq
An equivalent condition has been considered by Di\'osi and Kiefer
\cite{DioKie00} in a related context, but they did not make the
connection with physical realizability and measurements.

Equation (\ref{transprob}) implies that the quantum master
equation in the Wigner representation (\ref{OUE}) is equivalent to
a stochastic process for the mean displacements $\vec\mu$ of the
uniform Gaussian states with Wigner functions
$W(\vec{z};\vec{\mu})$. The  probability distribution for
$\vec{\mu}$ is governed by
 \beq \label{FPE}
\dot{p}(\vec{\mu}) =  \ro{ \vec{\del}^{T}{\bf K} \vec{\mu} + \half
\vec{\del}^{T} {\bf B}_{\bf V} \vec{\del} } p(\vec{\mu}),
\eeq
where here $\del_{m}=\dbd{\mu_{m}}$. This is analogous to \erf{clasme}.

To establish that condition (\ref{cond2}) is also satisfied we
thus have to show that  the stationary solution of \erf{FPE} is
$p(\vec\mu) = \wp(\vec\mu)$. From \erf{FPE} we get \cite{Gar85}
\beq
p_{\rm ss}(\vec{\mu}) = G(\vec{\mu};\vec{0},{\bf U}),
\eeq
where ${\bf U}$ is defined by
$
{\bf K}{\bf U} + {\bf U}{\bf K}^{T}  = {\bf B}_{\bf V}
$.
Now using Eqs.~(\ref{defVinf}) and (\ref{defB}) we obtain
\beq
{\bf U} = {\bf V}_{\rm ss} - {\bf V}.
\eeq
From \erf{defwp} it follows that $p_{\rm
ss}(\vec{\mu})=\wp(\vec{\mu})$ as desired.

\section{Example: The Atom Laser}
We now apply the above formalism to the problem of an atom laser.
We consider only a CW atom laser (which has not yet been
realized), which would consist of a continuously damped and
replenished Bose-Einstein condensate. The damping at rate $\kappa$
would produce a continuous beam of coherent atoms \cite{Wis97}.
The simplest quantum model for such a device uses a single mode
description of the condensate, with annihilation operator $\hat{a}
= (\hat{x}+i\hat{p})/2$. Assuming a Poissonian pump with rate
$\kappa\mu$, the stationary state of the laser is
\cite{Wis97,WalMil94}
\beq
\hat\rho_{\rm ss} = \int \frac{d\phi}{2\pi}
\ket{\sqrt{\mu}e^{i\phi}}\bra{\sqrt{\mu}e^{i\phi}}
= \sum_{n=0}^{\infty}
e^{-\mu}\frac{\mu^{n}}{n!}\ket{n}\bra{n} ,\label{mixnum}\label{mixcoh}
\eeq
Here $\ket{\sqrt{\mu}e^{i\phi}}$ is a coherent state (eigenstate of
$\hat{a}$) \cite{WalMil94}.

Equation (\ref{mixcoh}) shows two of the infinitude of different
ensemble representations of $\hat\rho_{\rm ss}$. The number state
ensemble $\cu{\ket{n}}_{n}$ is clearly PR by making continual
measurements of the environment in the atom number basis.
On the other hand, to determine the status of
the coherent state ensemble
$\cu{\ket{\sqrt{\mu}e^{i\phi}}}_{\phi}$, it is
necessary to examine the dynamics of the atom laser.

The simplest reasonable model for an atom laser comes from taking
the standard ideal optical laser model \cite{Wis97,Lou73} and
adding the interaction between condensate atoms \cite{WisVac01a},
governed by the Hamiltonian 
$(\hbar \kappa/4\mu) \chi \hat{a}\dg{}^{2} \hat{a}^{2}$, with
\beq
\chi = \frac{8\pi \mu \hbar a_{s} }{\kappa m}\int d^{3}{\bf r}|
          \psi({\bf r})|^{4}.
   \label{selfenergy}
\eeq
Here $\psi({\bf r})$ is the wavefunction for the condensate mode,
and $a_{s}$ is the $s$-wave scattering length. This nonlinear
Hamiltonian is difficult to treat exactly, so we linearize the
Lindbladian evolution about a mean field $\an{\hat{a}}=\sqrt{\mu}$
(as appropriate for considering the physical realizability of the
coherent state). We have shown elsewhere \cite{WisVac01a} that
this results in linear quantum dynamics for the Wigner function
$W(x,p)$ as in \erf{OUE} with
\beq
{\bf K} = \kappa \ro{
\begin{array}{cc}
    1 & 0  \\
    \chi & 0
\end{array}}\;;\;\;
{\bf D} = \kappa \ro{
\begin{array}{cc}
    2 & 0   \\
    0 & 2 + \nu
\end{array}}.
\eeq
Here $x$ and $p$ represent amplitude and phase fluctuations
respectively, and $\nu \geq 0$ is the excess phase diffusion.

To check if the coherent state ensemble is PR we simply need to
check the condition (\ref{defB}). We find
\beq
{\bf V} =  \ro{
\begin{array}{cc}
    1 & 0  \\
    0 & 1
\end{array}} \; \implies \;
{\bf B}_{\bf V} = \kappa\ro{
\begin{array}{cc}
    0 & -\chi   \\
    -\chi & 2 + \nu
\end{array}}.
\eeq
The matrix ${\bf B}_{\bf V}$ is positive semidefinite only for
$\chi =0$. That is, the coherent state ensemble is not PR for any
finite atomic interaction energy.

\section{Discussion}
In this letter we have introduced the necessary and sufficient
conditions for an ensemble of pure states to be physically
realizable (PR) as a stationary description of a Markovian system
in the long-time limit. Here physically realizable means that, by
measuring the environment of the system, an experimenter could
collapse the system state into a stochastically evolving pure
state, such that the proportion of time it is in a particular pure
state is equal to the weight of that pure state in the ensemble.
For uniform Gaussian ensembles for systems with linear dynamics we
derive a simple inequality to distinguish PR ensembles from non-PR
ensembles. The existence in general of these non-empty classes of
ensembles constitutes the {\em preferred ensemble fact}.

To illustrate this fact, we have shown that in a simple model for
an atom laser, the number state ensemble is preferred over the
coherent state ensemble because only the former is PR. This is due
to the presence of atomic interactions, described by a Hamiltonian
proportional to the $a\dg{}^{2}a^{2}$. We proved this using a
linearized analysis which included the loss and gain processes.
However, the result is perhaps not surprising since it is well
known in quantum optics that this Hamiltonian will turn a coherent
state into a squeezed state \cite{WalMil94}, which cannot be
described by mixture of coherent states.

This result is interesting because the ignorance interpretation of
the coherent state representation (``the laser is really in a
coherent state, I just haven't been bothered to find out which one
it is'') is very commonly adopted in quantum optics (see
Ref.~\cite{Mol96} for a discussion). Although it is tenable in
principle for most optical lasers where $\chi \ll 1$, it is not
tenable for atom lasers where we expect $\chi \gg 1$
\cite{WisVac01a}. So although atom lasers may give a coherent
output, the condensate cannot be meaningfully considered to be in
a coherent state. The relation between output coherence, and the
physical realizability of states of the condensate, will be
explored elsewhere.

HMW gratefully acknowledges conversations with T. Rudolph and R. 
Spekkens.

\end{multicols}


\begin{references}
\bibitem{fn1}
The time scale for the measurement must be much longer than the
correlation time of the environment, but much smaller than any
relevant system dynamical time scale.
\end{references}
\end{document}